\documentstyle[12pt]{article}

\begin{document}

\title{\bf Distribution of Forces in Gravitationally Clustered Systems}
\author{ Eliani {\sc Ardi}$^{1,2}$ and Shogo {\sc Inagaki}$^1$}
\date{}
\maketitle
\centerline {$^1${\it Department of Astronomy, Faculty of Science, Kyoto University, Sakyo-ku, Kyoto 606-01}}
\centerline {$^2${\it Department of Astronomy, Faculty of Science, Institute of Technology Bandung, }}
\centerline {\it Jl. Ganesha 10, Bandung 40132, Indonesia}
\centerline {\it E-mail(EA): ardi@kusastro.kyoto-u.ac.jp}
\vspace{0.2 in}
\centerline {(Received 1996 September 20 ; Accepted 1996 December 9)}
\vspace{0.5 in}

\begin{flushleft}
{\bf Abstract}
\end{flushleft}

We have studied the distribution of forces in gravitational systems through numerical experiments. Data were taken from an $\it N$-body simulation in an expanding universe. Before clustering, the distribution of random forces was represented as a Holtsmark distribution; the nearest-neighbor distribution is also shown as a comparison. The analytical and simulation distributions are in good agreement. When clustering becomes strong, the simulation result showed that the contribution of the forces acting on each galaxy, which is generated from all other galaxies, is almost entirely due to the gravitational attraction of its nearest neighbor. This implies that nearest-neighbor galactic encounters may play the main role in the dynamics of galaxy clustering.

\begin{flushleft}
{\bf Key words:}
\end{flushleft}
Cosmology --- Galaxies: clusters of --- Gravitation --- Numerical methods 
\newpage

\begin{flushleft}
{\bf 1. Introduction}
\end{flushleft}

One of the fundamental problems in galactic dynamics is to analyze of the nature of the forces acting on a galaxy that is a member of a system of galaxies in the large-scale structure of the universe. The average force acting on each galaxy in gravitational systems, such as clusters of galaxies, can be naturally decomposed as

%%%%%%%%%% Equation 1 %%%%%%%%%%
\begin{equation}
   \mbox{\boldmath $\Upsilon$} = \mbox{\boldmath $K$} + \mbox{\boldmath $F$}.  
\end{equation}
%%%%%%%%%%%%%%%%%%%%%%%%%%%%%%%% 

Here, {\boldmath $K$} is a slowly varying part, derivable from a potential function, which is due to the smoothed-out distribution of the galaxies in the system. The random force, {\boldmath $F$}, is due to a relatively rapid statistical fluctuation of the immediate local neighborhood. As a consequence of fluctuations, the actual value of {\boldmath $F$} at any particular moment depends on the instantaneous complexion of the local distribution of galaxies. In this sense {\boldmath $F$} is a stochastic force, which can be described through the assignment of a force-distribution function $P$({\boldmath $F$}), just like any stochastic variable.
The probability of a stationary force distribution in gravitational systems was first given by Chandrasekhar and von Neumann (1942) for an infinite homogeneous system. It includes two basic assumptions. First, the stars surrounding the test star on which the force is being calculated are distributed with a uniform probability density. This implies a constant density in the local neighborhood, with no correlations among the positions of the stars. Second, it allows the number of stars to tend to infinity while keeping the density constant. This force-probability distribution is given by the Holtsmark distribution. 

Under the above assumptions, Chandrasekhar (1943) obtained a distribution law of the nearest neighbors in a random distribution of stars. It denotes the probability of force which comes from the nearest-neighbor stars, which is called the nearest-neighbor distribution. Chandrasekhar also showed analytically that for $|${\boldmath $F$}$|\rightarrow \infty$ the nearest-neighbor distribution is in exact agreement with the Holtsmark distribution. The physical meaning of this agreement is that the dominant contribution to {\boldmath $F$} is made by the nearest-neighbor galaxies.

Attemps to examine the probability of the force distribution in an inhomogeneous system were begun by Kandrup (1980, 1981). He showed that the distribution of random force in an inhomogeneous system is a generalization of the Holtsmark distribution.
Antonuccio-Delogu and Atrio-Barandela (1992) derived the force distribution in weakly clustered systems, where the two-point correlation function $\xi(r)$ is $<<$ 1.

A numerical experiment performed by Del Popolo (1996) shows that the analytical results of Kandrup (1980) and Antonuccio-Delogu and Atrio-Barandela (1992) give a good description of the random forces in inhomogeneous and weakly clustered systems.

In the present work, we considered the evolution of the force distribution function which comes from all the galaxies in the system (hereafter Total-Force Distribution), and from the nearest galaxies (hereafter Nearest-Neighbor Distribution) before and during gravitational clustering of galaxies in an expanding universe, through a numerical experiment. The motivation of searching for the contribution of the force acting from the nearest galaxies is to examine the influence of the nearest-neighbor galactic encounters in gravitationally clustered systems.     

Before clustering, it was assumed that galaxies are distributed under a Poisson distribution. In this case, we make comparisons from both the analytical and simulation results. Analytically, we applied the Holtsmark law and the nearest-neighbor law. During clustering,  all 1000 particles, each of which represents a galaxy, are examined for a $\Omega = 1.0$ universe (Peebles, 1980) for timescales of $a/a_o$ = 4.0, 8.0, 16.0, 22.63, 32.0, where $a$ is the scale factor. 

\begin{flushleft}
{\bf 2. Total-Force Distribution and Nearest-Neighbor Distribution}
\end{flushleft}

The stochastic force {\boldmath $F$} was obtained by subtracting the mean force from the total force at the $i$-th galaxy,

%%%%%%%%%% Equation 2 %%%%%%%%%%
\begin{equation}
     \mbox{\boldmath $F$}_{i} = -G \sum_{j \neq i}^N \frac {m}{|\mbox{\boldmath $r$}_{ij}|^3} \mbox{\boldmath $r$}_{ij} + \frac {G M_{tot}\mbox{\boldmath $r$}_i}{R^3},
\end{equation}
%%%%%%%%%%%%%%%%%%%%%%%%%%%%%%%%
where $m$ denotes the mass of the field galaxy, {\boldmath $r$}$_{ij}$ is the position of the $i$-th galaxy relative to the $j$-th galaxy, {\boldmath $r$}$_{\rm i}$ is the position of the $i$-th galaxy to the origin, $N$ denotes the number of neighboring galaxies, and $G$ is the gravitational constant. $M_{\rm tot}$ is the total mass of the system and $R$ is its radius. Since the value of {\boldmath $F$} at a given time depends on the instantaneous  positions of all other galaxies, it is subject to fluctuations as these positions change. Even if the galactic distribution is stationary and homogeneous, {\boldmath $\Upsilon$} fluctuates around an average value due to local Poisson fluctuation in the number density of neighboring galaxies. The fluctuating part of {\boldmath $\Upsilon$} that was indicated as {\boldmath $F$} can be represented by the probability distribution of the stochastic force method $P$({\boldmath $F$}). The probability for the occurence of {\boldmath $F$} in the range {\boldmath $F$}+d{\boldmath $F$} is
%%%%%%%%%% Equation 3 %%%%%%%%%%
\begin{equation}
     P(\mbox{\boldmath $F$})dF_{x}dF_{y}dF_{z} = P(\mbox{\boldmath $F$})d\mbox{\boldmath $F$}.
\end{equation}
%%%%%%%%%%%%%%%%%%%%%%%%%%%%%%%%
An explicit form of this distribution is shown as the Holtsmark distribution $P$({\boldmath $F$}) (Chandrasekhar, von Neumann 1942),
%%%%%%%%%% Equation 4 %%%%%%%%%%
\begin{equation}
     P(|\mbox{\boldmath $F$}|) = \frac {2 Q}{\pi |\mbox{\boldmath $F$}|} \int_0^\infty \exp[- (x Q / |\mbox{\boldmath $F$}|)^{3/2}] x \sin x dx,
\end{equation}
%%%%%%%%%%%%%%%%%%%%%%%%%%%%%%%%
where the normal field $Q$ is defined by
%%%%%%%%%% Equation 5 %%%%%%%%%%
\begin{equation}
     Q = 2.6031 G (m^{3/2} n)^{2/3}
\end{equation}
%%%%%%%%%%%%%%%%%%%%%%%%%%%%%%%%
($n$ denotes the average number of particles per unit volume).
This equation was applied to describe the force distribution analytically in the initial condition when galaxies are randomly distributed. In particular, the value of $|${\boldmath $F$}$|$ that has the maximum probability of occurence is found to be $\sim$ 1.6 $Q$.
The asymptotic behavior of the distribution function in a strong field is found by
%%%%%%%%%% Equation 6 %%%%%%%%%%
\begin{equation}
     P(|\mbox{\boldmath $F$}|) \simeq 2 \pi G^{3/2} m^{3/2} n |\mbox{\boldmath $F$}|^{-5/2} \ (|\mbox{\boldmath $F$}| \rightarrow \infty).
\end{equation}
%%%%%%%%%%%%%%%%%%%%%%%%%%%%%%%%

The law of distribution of the nearest neighbor is given by
%%%%%%%%%% Equation 7 %%%%%%%%%%
\begin{equation}
     P(r)dr = \exp (-4 \pi r^3 n/3)4 \pi r^2 n dr.  
\end{equation}
%%%%%%%%%%%%%%%%%%%%%%%%%%%%%%%%
$P(r)dr$ denotes the probability that the nearest neighbor to a particle occurs between $r$ and $r + dr$.
Since in a first-neighbor approximation
%%%%%%%%%% Equation 8 %%%%%%%%%%
\begin{equation}
     |\mbox{\boldmath $F$}| = Gm/r^2,
\end{equation}
%%%%%%%%%%%%%%%%%%%%%%%%%%%%%%%%
we readily obtain 
%%%%%%%%%% Equation 9 %%%%%%%%%%
\begin{eqnarray}
     P(|\mbox{\boldmath $F$}|)d|\mbox{\boldmath $F$}| =
     \exp [-4 \pi (Gm)^{3/2}n/3 |\mbox{\boldmath $F$}|^{3/2}] 2 \pi (Gm)^{3/2}
     n|\mbox{\boldmath $F$}|^{5/2} d|\mbox{\boldmath $F$}|,
\end{eqnarray}
%%%%%%%%%%%%%%%%%%%%%%%%%%%%%%%%
which has an asymptotic behavior at strong fields, as
%%%%%%%%%% Equation 10 %%%%%%%%%%
\begin{equation}
     P(|\mbox{\boldmath $F$}|) \simeq 2 \pi (Gm)^{3/2}n |\mbox{\boldmath $F$}|^{-5/2} \ (|\mbox{\boldmath $F$}| \rightarrow \infty).
\end{equation}
%%%%%%%%%%%%%%%%%%%%%%%%%%%%%%%
It can be seen to be in exact agreement with formula (6), derived from the Holtsmark distribution. Analytically, we can compare both the Holtsmark distribution and the nearest-neighbor distribution by using equations (4) and (9 ). 

A simulation of the total-force distribution can be obtained by using
%%%%%%%%%% Equation 11 %%%%%%%%%%
\begin{eqnarray}
     |\mbox{\boldmath $F$}_j| = \Bigl [\ \Bigl (N x_j -\sum {i \neq j}^N \frac {x_j - x_i}
     {|x_j - x_i|^3} \Bigl )^2 + 
     \Bigl (N y_j -  \sum_{i \neq j}^N \frac {y_j - y_i}{|y_j - y_i|^3} 
     \Bigl )^2 + \Bigl (N z_j -\sum_{i \neq j}^N \frac {z_j - z_i}{|z_j - z_i|^3}+ \Bigl )^2\ \Bigl ]^{1/2}.
\end{eqnarray}
%%%%%%%%%%%%%%%%%%%%%%%%%%%%%%%
Here, we use the assumption that $G$ = $m$ = $R$ = 1. $N$ is the number of galaxies, {\boldmath $F$}$_j$ is the stochastic force acting on the $j$-th galaxy, which was generated by all of its neighbors. The nearest-neighbor distribution was simulated by using  
%%%%%%%%%% Equation 12 %%%%%%%%%%
\begin{eqnarray}
     |\mbox{\boldmath $F$}_j| = \Bigl [\ \Bigl ( \frac {x_j - x_n}{|x_j - x_n|^3} \Bigl )^2 + \Bigl ( \frac {y_j - y_n}{|y_j - y_n|^3} \Bigl )^2 + \Bigl ( \frac {z_j - z_n}{|z_j - z_n|^3} \Bigl )^2\  \Bigl ]^{1/2},
\end{eqnarray}
%%%%%%%%%%%%%%%%%%%%%%%%%%%%%%%
where $(x_n,y_n,z_n)$ is the coordinate of the nearest neighbor of the $j$-th galaxy.

\newpage

\begin{flushleft}
{\bf 3. Results}
\end{flushleft}

The distributions of the forces before clustering are presented in figure 1 and figure 2. They show the Holtsmark distribution as well as the nearest-neighbor distribution for both the analytical curve and the simulation result. The Holtsmark distribution was produced by a stronger force than that in the nearest-neighbor distribution. In the simulation, 1000 particles were initially randomly distributed inside a sphere of radius $R$ = 1. Each particle represented a galaxy. We can see that the simulation result gives good agreement with the analytical one. From this result we see that we may inspect the distribution of forces in clustered states with 1000 particles.

\begin{center}
%\hspace{1 in}
------------~~~~~~~~------------ \\
%\hspace{1 in}
~Figure 1~~~~~~~~  Figure 2 \\
%\hspace{1 in}
------------~~~~~~~~------------ \\
\end{center}

Figure 3 shows projected distributions of galaxies during clustering for six timescales of $a / a_o$, where $a$ is the cosmological expansion parameter with an initial value of $a_o$. These were all projected views onto the $X$--$Y$ plane. The maximum comoving radius of the simulation is always scaled to $R = 1$.

\begin{center}
------------

Figure 3

------------
\end{center}

Figures 4 -- 9 present the evolution of the simulation result of both the total-force distribution and nearest-neighbor distribution. We used equations (11) and (12) for a simulation of the force distribution, which was applied for $N$=1000 particles. For each particular value of $a / a_o$, we notice that both distributions are very similar in regions of strong force, except in regions of weak force. This demonstrates that if a galaxy experiences large forces, they are almost certainly due to the single nearest-neighbor galaxy. The region in which the two distributions differ most markedly is when the force is very weak. The frequency of occurence of the nearest-neighbor distribution within the weak force is higher than that of the total force distribution. 

Part b of figures 4 -- 9 illustrates the evolution of the distribution in a strong-force region. As the value of $a/a_o$ becomes bigger, the total-force distribution becomes more similar to the nearest-neighbor distribution. After several time scales of expansions, when clustering is strong, the contribution of the force generated from the nearest neighbor of the galaxy becomes stronger. The force acting on the galaxies per unit mass is almost entirely due to the gravitational attraction of the nearest neighbors. It is also noted that the frequencies of occurrence of forces are nearly constant over a wide range of the force strength.

\begin{center}
------------

Figure 4

------------
\end{center}

\begin{center}
------------

Figure 5

------------
\end{center}

\begin{center}
------------

Figure 6

------------
\end{center}

\begin{center}
------------

Figure 7

------------
\end{center}

\begin{center}
------------

Figure 8

------------
\end{center}

\begin{center}
------------

Figure 9

------------
\end{center}

\begin{flushleft}
{\bf 4. Conclusions}
\end{flushleft}

The result of a simulation of the force distribution before clustering is in good agreement with the analytical distribution. Therefore, we can use the data from an $N$-body simulation for examining the evolution of the nearest-neighbor and total-force distribution, although there is as yet no analytical function with which to make a comparison.

In order to follow the influence of galactic encounters, we examine the evolution of a nearest-neighbor distribution, which is compared with the total-force distribution. Figures 4 -- 9 show that as the evolution of clustering is in progress, the total-force distribution and the nearest-neighbor distribution agree over most of the range of {\boldmath $F$}, especially for stronger fields. The physical meaning of this result is that the force acting on galaxies per unit mass is almost entirely due to the gravitational attraction of their nearest neighbors.

We conclude that the influence of the nearest-neighbor galactic encounters becomes dominant when clustering is strong. The force acting on each galaxy in gravitationally clustered systems in the expanding universe is almost entirely due to the gravitational attraction of the nearest neighbor. Nearest-neighbor galactic encounters might well play the main role in the dynamics of galaxy clustering. If so, this may have two implications. First, in a gravitationally clustered system, such as clusters of galaxies, each encounter could be treated as a two-body encounter representing the perturber galaxy and its nearest-neighbor galaxy. Second, the collisionless $N$-body simulation method (Binney, Tremaine 1987) may not be well suited for studying the dynamics of galaxy clustering, in which nearest-neighbor galactic encounters play the main role, because this method assumes that the force from the nearest neighbor is negligible and the collective force is dominant.
\vspace{12 pt}

We are grateful to Dr.T.Tsuchiya for reading the manuscript and making useful comments. E.A. thanks all members of Department of Astronomy, University of Kyoto, for their hospitality during her stay. She is indebted to the Ministry of Education, Science, Sport and Culture of Japan for financial support.

\begin{flushleft}
{\bf References}
\end{flushleft}

\begin{description}
\item
Antonuccio-Delogu V., Atrio-Barandela F. 1992, ApJ 392, 403
\item
Binney J., Tremaine S. 1987, Galactic Dynamics (Pricenton University Press) pp90,91
\item
Chandrasekhar S. 1943, Rev. Mod. Phys. 15, 68 
\item
Chandrasekhar S., von Neumann J. 1942, ApJ 95, 489 
\item
Del Popolo A. 1996, A\&A 305, 999
\item
Kandrup H.E. 1980, Phys. Rep. 63, 1
\item
Kandrup H.E. 1981, ApJ 244, 1039
\item
Peebles P.J.E. 1980, The Large Scale Structure of The Universe (Pricenton University Press, Pricenton) pp51,52
\item
\end{description}

\newpage

\begin{flushleft}
{\bf Figure Captions}
\end{flushleft}

Fig. 1. Holtsmark distribution in the analytical curve and simulation result.

Fig. 2. Nearest-neighbor distribution in the analytical curve and simulation result.

Fig. 3. Six projected distributions for time scales of $a / a_o$ = 1.0, 4.0, 8.0, 16.0, 22.63, 32.0.

Fig. 4. Total force distribution (solid line) and nearest-neighbor distribution (short dashes line) at $a / a_o$ = 1.0 (a) in whole; (b) enlargement of the large-force region.

Fig. 5. Total force distribution (solid line) and nearest-neighbor distribution (short dashes line) at $a / a_o$ = 4.0 (a) in whole; (b) enlargement of the large-force region.

Fig. 6. Total force distribution (solid line) and nearest-neighbor distribution (short dashes line) at $a / a_o$ = 8.0 (a) in whole; (b) enlargement of the large-force region.

Fig. 7. Total force distribution (solid line) and nearest-neighbor distribution (short dashes line) at $a / a_o$ = 16.0 (a) in whole; (b) enlargement of the large-force region.

Fig. 8. Total force distribution (solid line) and nearest-neighbor distribution (short dashes line) at $a / a_o$ = 22.63 (a) in whole; (b) enlargement of the large-force region.

Fig. 9. Total force distribution (solid line) and nearest-neighbor distribution (short dashes line) at $a / a_o$ = 32.0 (a) in whole; (b) enlargement of the large-force region.

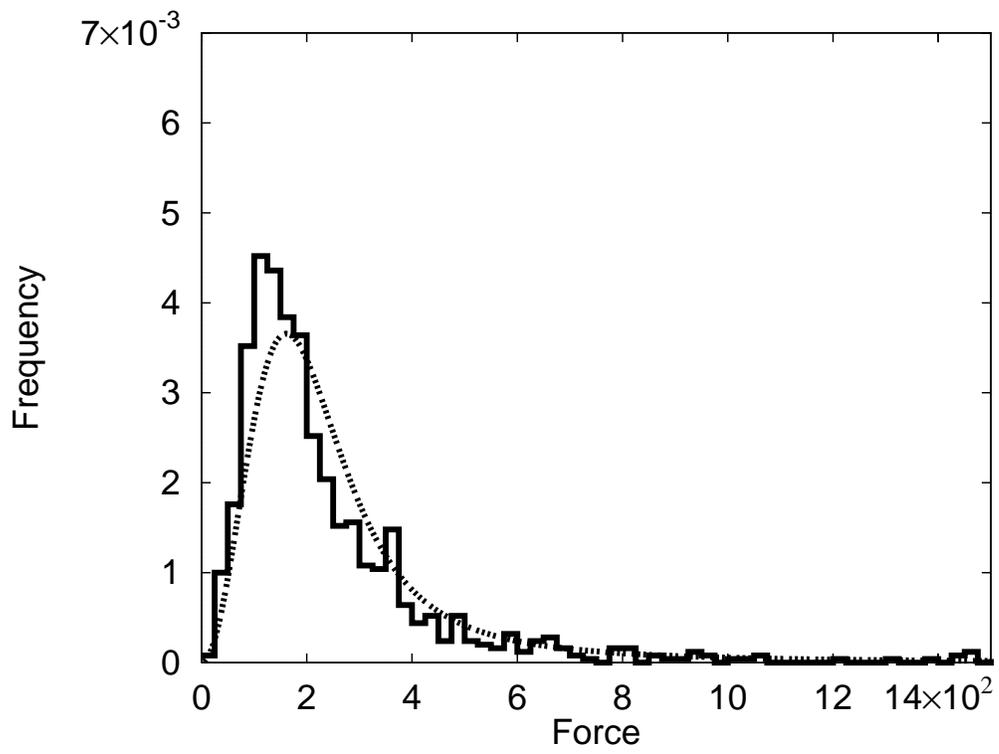
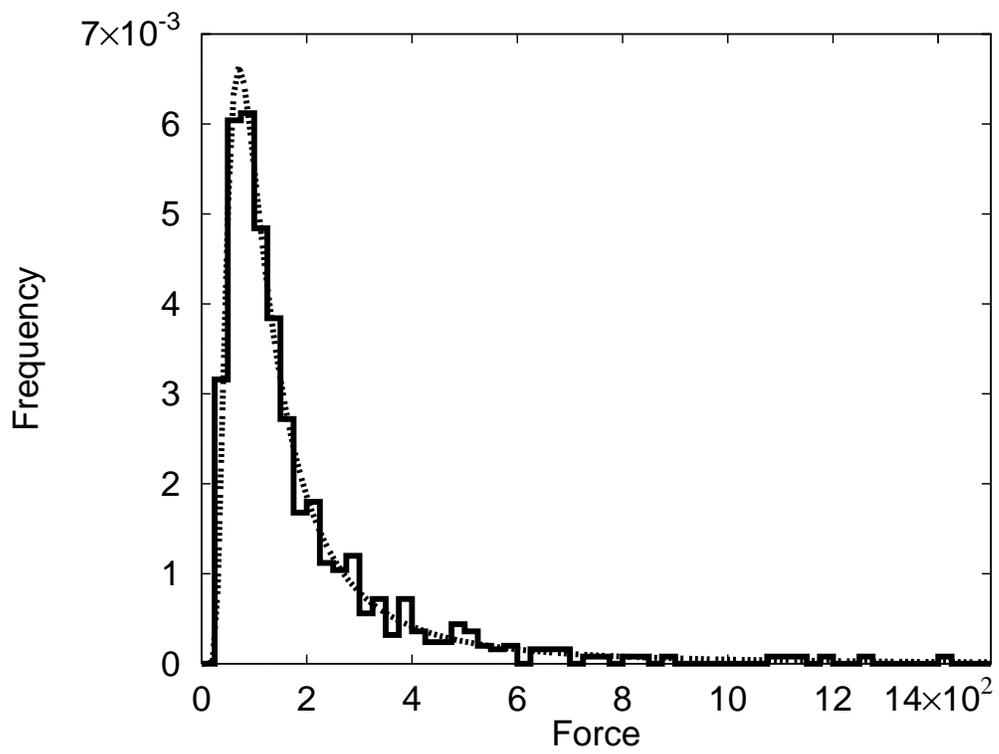
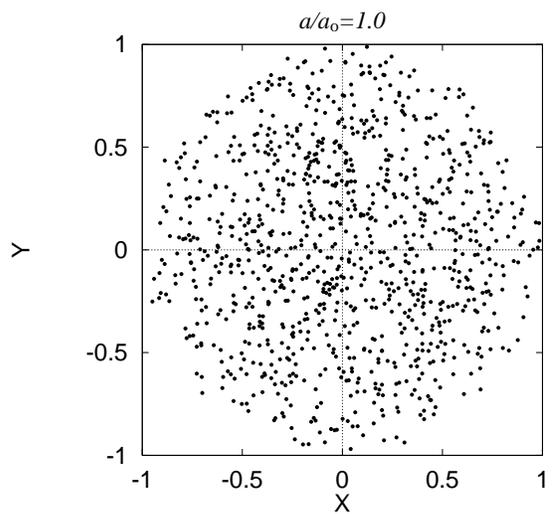
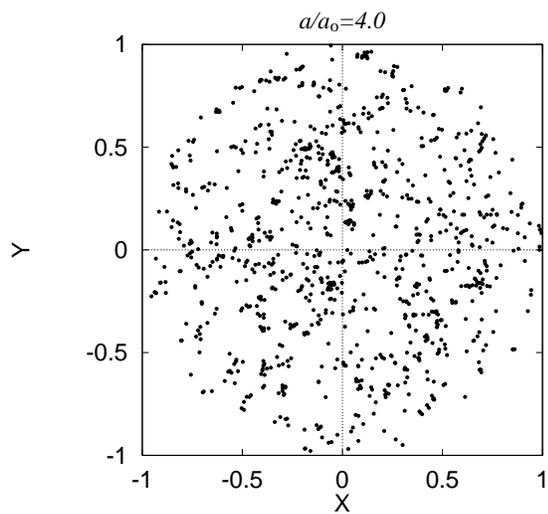
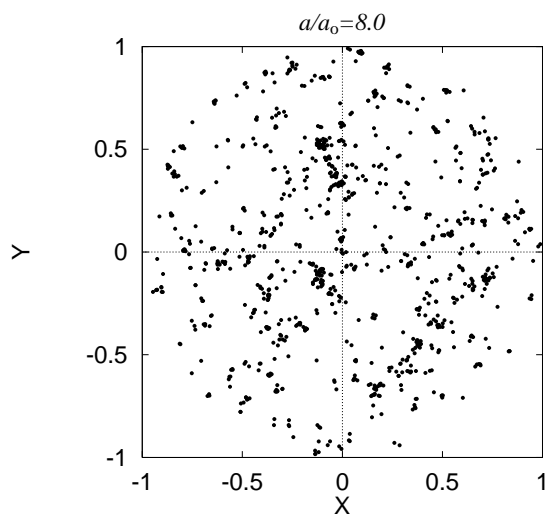
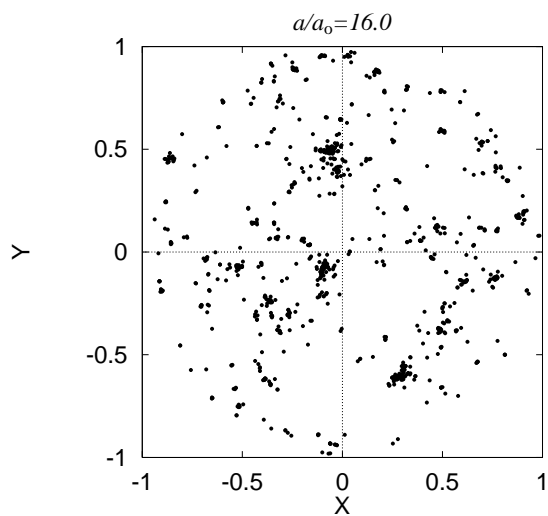
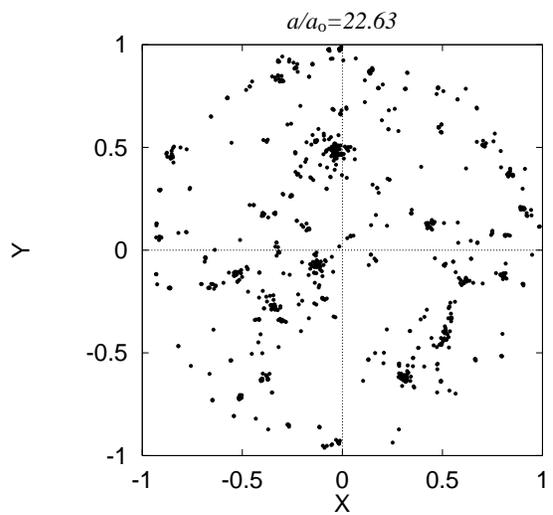
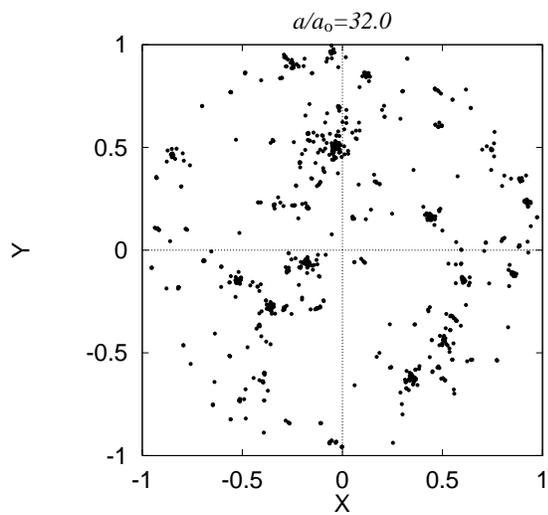
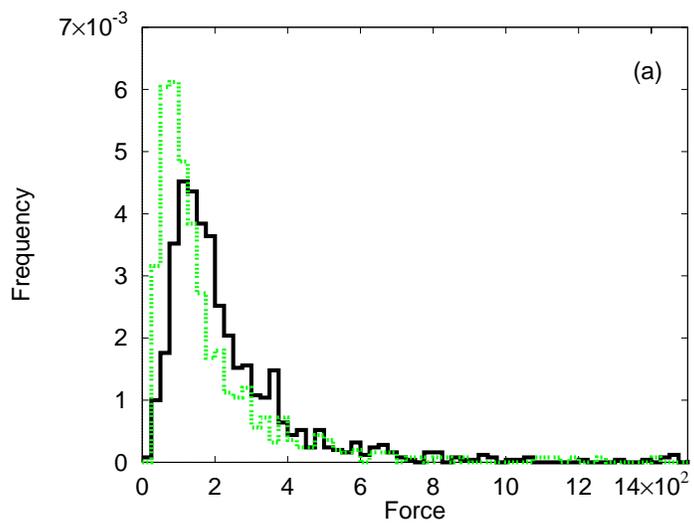
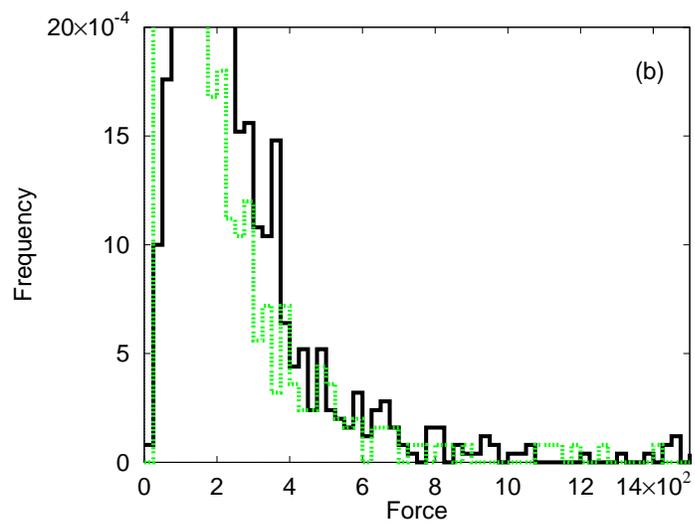
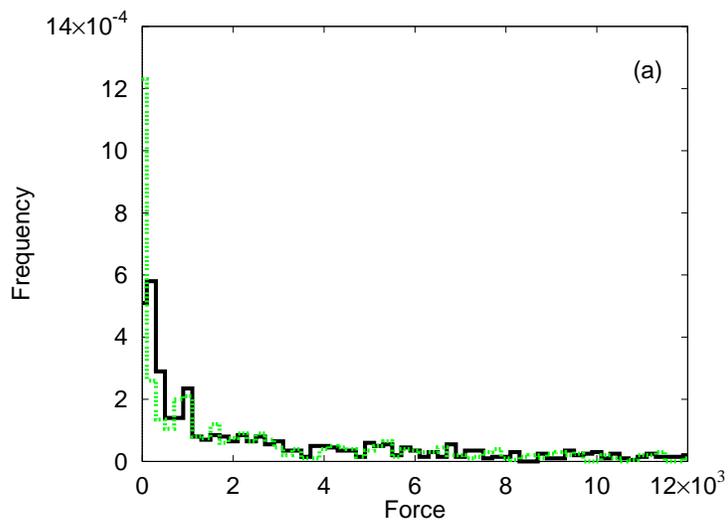
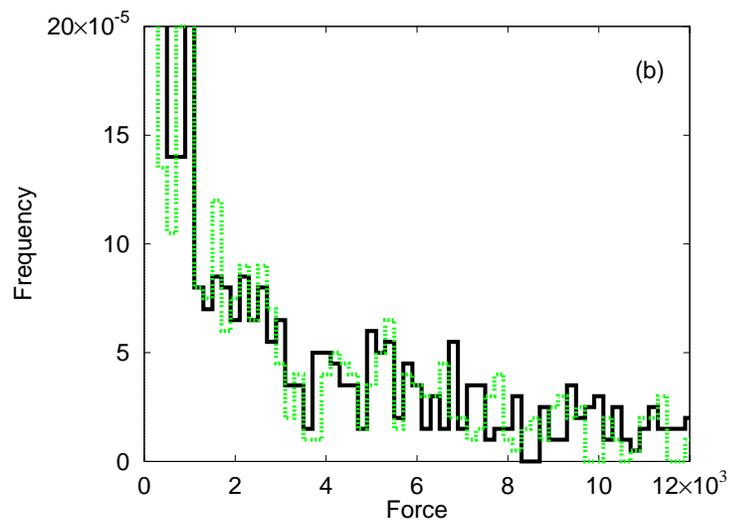
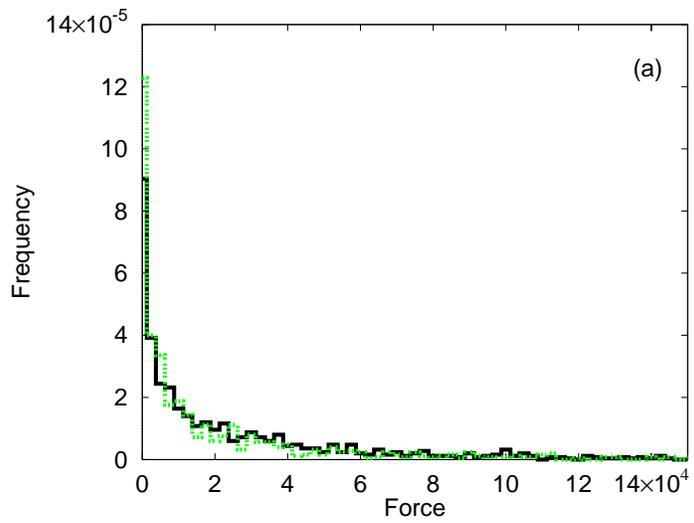
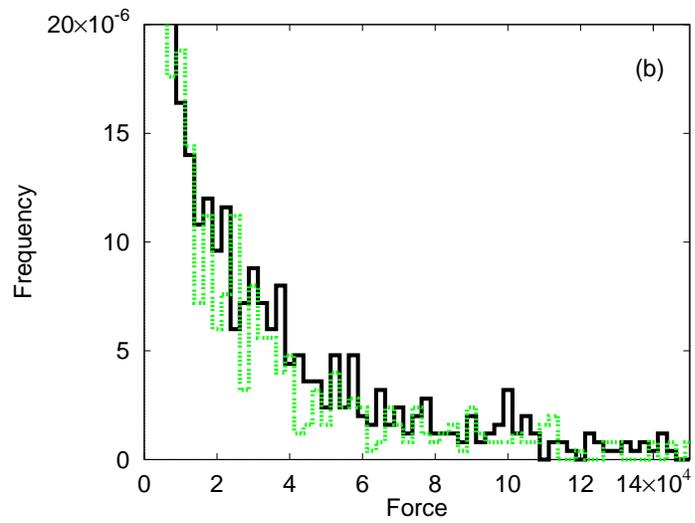
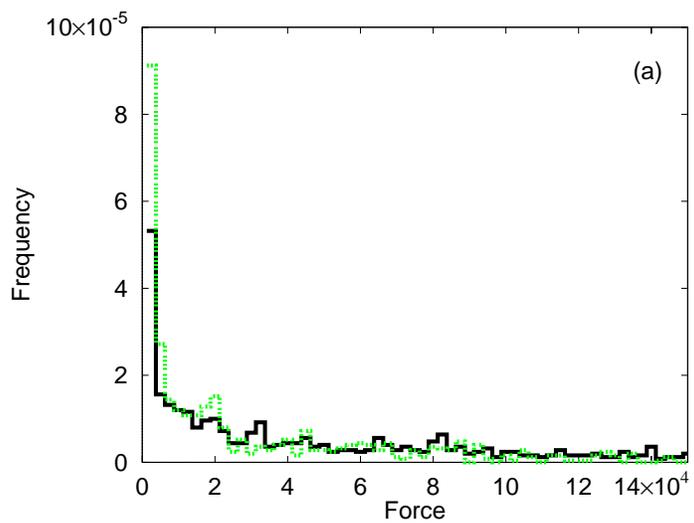
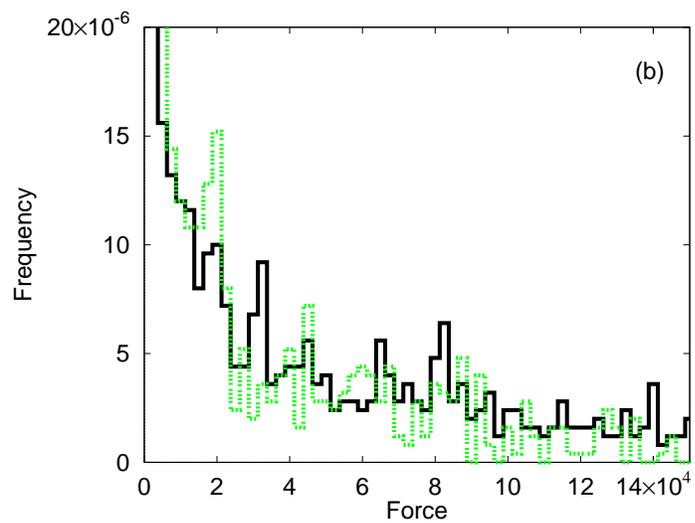
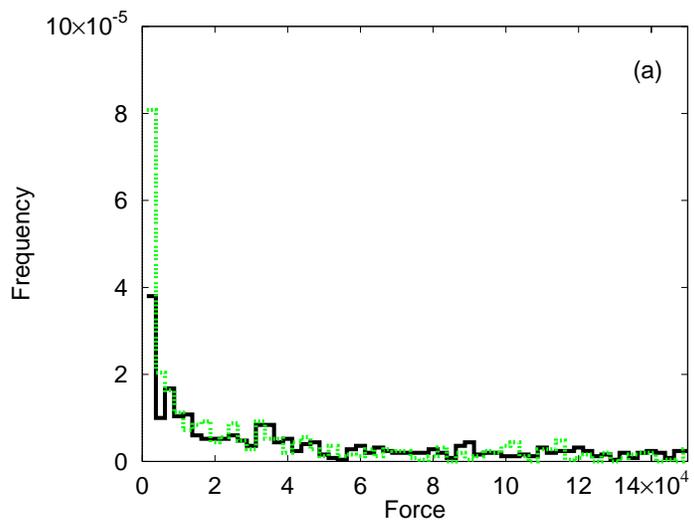
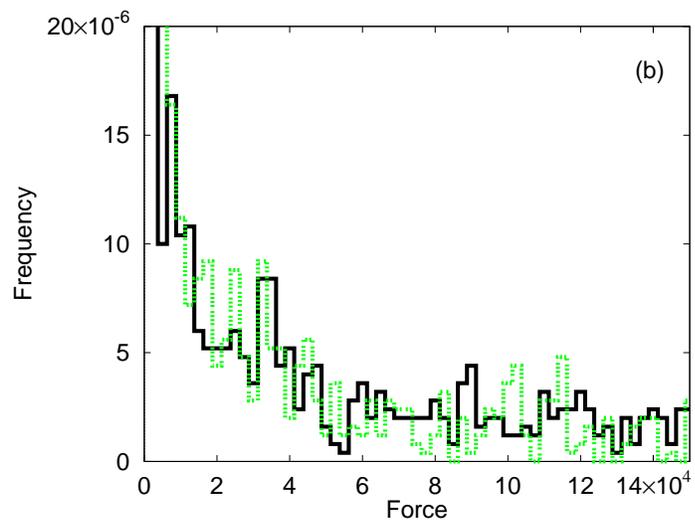
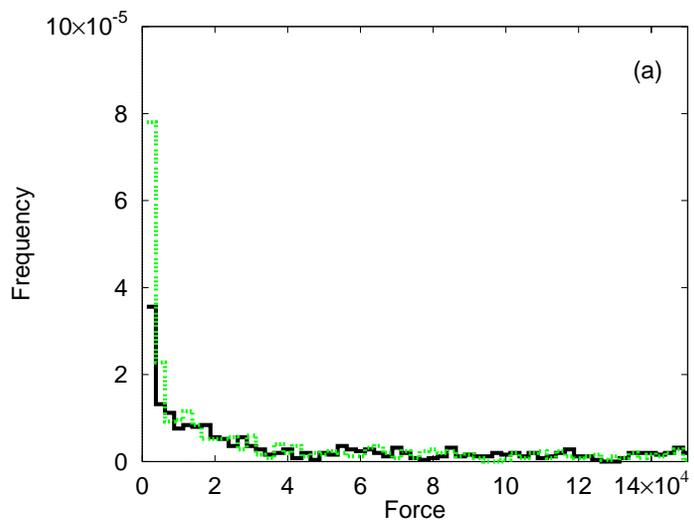
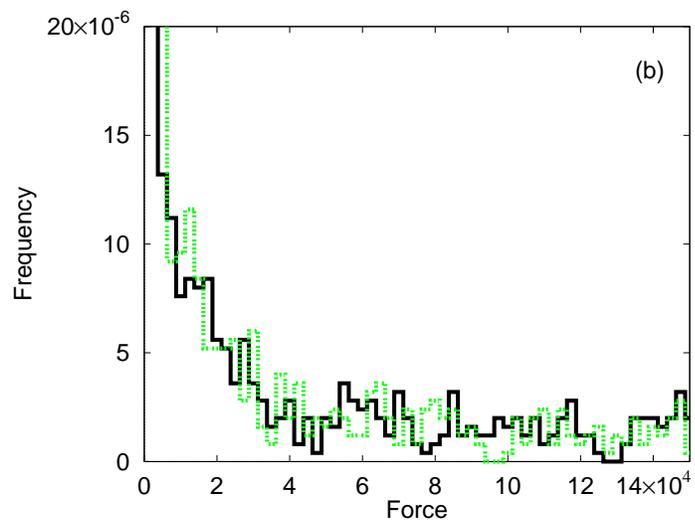
\end{document}